\documentclass[12pt]{article}
\pdfoutput=1
\usepackage[colorlinks=true,linkcolor=blue,urlcolor=blue,
  citecolor=blue]{hyperref}
\usepackage{amsfonts}
\usepackage{color}
\usepackage{fullpage}
\usepackage{amssymb,graphicx}
\usepackage{amsmath}
\usepackage[margin=0.5cm]{caption}
\usepackage{hyperref} 
\usepackage{cite}
\usepackage{upgreek}

\def\l{\left(}
\def\r{\right)}

\newcommand{\be}{\begin{equation}}
\newcommand{\ee}{\end{equation}}
\newcommand{\bea}{\begin{eqnarray}}
\newcommand{\eea}{\end{eqnarray}}
\newcommand{\bg}{\begin{gather}}
\newcommand{\eg}{\end{gather}}
\newcommand{\bseq}{\begin{subequations}}
\newcommand{\eseq}{\end{subequations}}

\begin{document}
\baselineskip=15.5pt
\begin{titlepage}
\begin{center}
{\Large\bf Heating up the Galaxy with Hidden Photons }\\
\vspace{0.5cm}
{ \large
Sergei Dubovsky$^a$ and Guzm\'an Hern\'andez-Chifflet$^{a,b}$ 
}\\
\vspace{.45cm}
{\small  \textit{   $^a$Center for Cosmology and Particle Physics,\\ Department of Physics,
      New York University\\
      New York, NY, 10003, USA}}\\ 
\vspace{.25cm}      
      {\small  \textit{   $^b$Instituto de F\'isica, Facultad de Ingenier\'ia,\\ Universidad de la Rep\'ublica,\\
      Montevideo, 11300, Uruguay}}\\ 
\end{center}
\begin{center}
\begin{abstract}
We elaborate on the dynamics of  ionized interstellar medium in the presence of hidden photon dark matter. Our main focus is the ultra-light regime, where the hidden photon mass is smaller than the plasma frequency in the Milky Way. We point out  that as a result of the Galactic plasma shielding direct detection of ultra-light photons in this mass range is especially challenging.  However, we demonstrate that ultra-light 
hidden photon dark matter provides a powerful heating source for the ionized interstellar medium.
This results in 
 a strong bound on the kinetic mixing between hidden and  regular photons  all the way down to the hidden photon masses of order $10^{-20}$ eV. 
\end{abstract}
\end{center}
\end{titlepage}

\section{Introduction}

The nature of dark matter is one of the most exciting and pressing puzzles in particle physics and cosmology.  Some dark matter candidates (most notably, the QCD axion and weakly interacting massive particles) have a more solid theoretical motivation than the others. However,
given our current ignorance about composition and dynamics of the dark sector, a reasonable strategy in addressing the dark matter challenge is to leave no stone unturned, and explore as much consistent options and regions in the parameter space, as possible. 

One proposal in the off-the-beaten-path category, which attracted  considerable attention recently (see, e.g.,  \cite{Nelson:2011sf,arias2012wispy, an2013new,An:2014twa,Chaudhuri:2014dla,Graham:2015rva}), is that dark matter is made up of massive vector bosons, known as hidden photons. The simplest viable scenario for  hidden photon dark matter assumes that no additional light particles carrying a hidden electric charge are present. In particular, this implies that
the hidden photon mass term is of the St\"uckelberg type, {\it i.e.} no Higgs field is involved.
This is the setup we consider here. In addition to the hidden photon mass $m$, this minimal model is characterized by a single dimensionless parameter $\varepsilon$, which determines  the strength of kinetic mixing between the hidden and the ordinary photon \cite{Holdom:1985ag}.  It is convenient to work in the field basis such that the quadratic action is diagonal.
The kinetic mixing translates then into a direct coupling between the hidden photon and the Standard Model (SM) electric current $J^\mu$.
The resulting Lagrangian is 
\begin{equation}
\label{Lagr}
\mathcal{L} = -\frac{1}{4}F_{\mu\nu}F^{\mu\nu} -\frac{1}{4}\tilde{F}_{\mu\nu}\tilde{F}^{\mu\nu} +\frac{m^2}{2}\tilde{A}_{\mu}\tilde{A}^{\mu} - {e\over (1+\varepsilon^2)^{1/2}}J^\mu\left(A_\mu + \varepsilon\tilde{A}_\mu\right),
\end{equation} 	
where $A_\mu$ ($\tilde{A}_\mu$) and  ${F}_{\mu\nu}$ ($\tilde{F}_{\mu\nu}$) stand for the visible (hidden) photon gauge fields and field strengths respectively.  Our notations are different from the standard ones, resulting in a somewhat unconventional  factor of $(1+\varepsilon^2)^{-1/2}$ in (\ref{Lagr}). Clearly, the two conventions agree at $\varepsilon\ll 1$. The advantage of the one adopted here is that $e$ is equal to the physical electric charge, as measured at high energies
.

\begin{figure}[t!!]
	\centering
	\includegraphics[width=\textwidth]{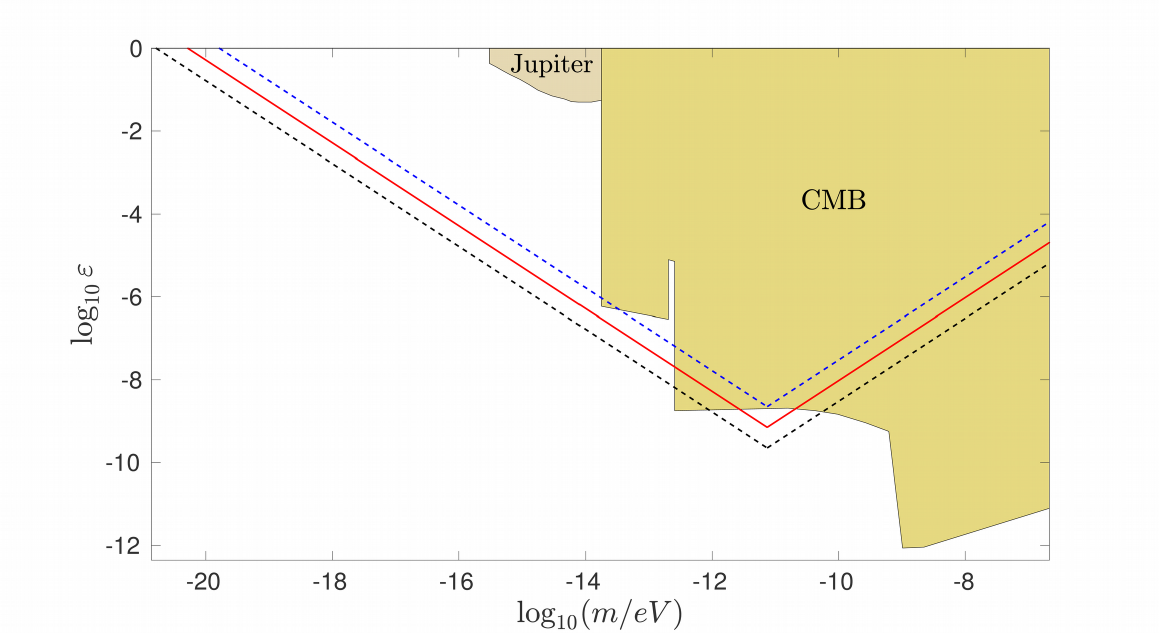}
	\caption{Bounds on the kinetic mixing $\varepsilon$ as a function of the dark matter hidden photon mass $m$ for three choices of parameters: black dashed lines corresponds to 
	the cooling rate $ Q\sim 10^{-27}$~erg/s $n_H$, with $n_H\sim 0.1$~cm$^{-3}$ (an ``optimistic" choice), red solid line to $ Q\sim 10^{-26}$~erg/s $n_H$, with $n_H\sim 0.1$~cm$^{-3}$ (a ``moderate" choice), and
	the blue dotted to $ Q\sim 10^{-26}$~erg/s $n_H$, with $n_H\sim 1$~cm$^{-3}$ (a ``conservative" choice) with $T_e = 8000 \text{ K}$, $n_e=0.04$~cm$^{-3}$ in all the cases.
The two filled regions show  the CMB bound \cite{arias2012wispy,Chaudhuri:2014dla} and the bound from the Jupiter magnetic field \cite{Jaeckel:2010ni}. This plot does not include constraints from black hole superradiance, which are briefly discussed in Section~\ref{sec:last}.}
		\label{fig_bound}
\end{figure}

The parameter space of this model has been constrained by a variety of laboratory experiments, astrophysical observations and cosmological arguments. In the present work we are interested in the ultra-light regime, $m\lesssim 10^{-11}$~eV.
This region of parameters is especially interesting from the dynamical view-point because the interstellar medium cannot be treated as an empty space, as can be seen from comparing the hidden photon mass to the plasma frequency  
\be
\label{wp}
\omega_p = \l\frac{4\pi n_e \alpha}{m_e}\r^{1/2}
\approx 7.4\cdot 10^{-12}\mbox{\rm ~eV}
\l {n_e\over 0.04
\mbox{\rm 
~cm}^{-3}
}\r^{1/2}\;,
\ee
where  $n_e\sim 0.04
\mbox{\rm 
 ~cm}^{-3}$ is a typical value of the free electron density in the vicinity of the Solar system. For many purposes  plasma frequency may  be thought of as an effective photon mass, so for $m<\omega_p$ matter effects make the conventional photon heavier than the hidden one. If the hidden photon were heavier than the SM one in the intergalactic space, a resonant conversion of hidden photons into SM photons would occur in the early Universe \cite{arias2012wispy}.
  Depending on the hidden photon mass, this may result either in cosmic microwave background (CMB) distortions or in the depletion of the amount of dark matter at low redshifts  compared to its abundance at recombination.
  This results in the tight constraint on the ultra-light photon dark matter presented in Fig.~\ref{fig_bound}.
   
 For $m\lesssim 10^{-14}$~eV the hidden photon is lighter than the SM one even in the intergalactic space, 
 so the resonant condition is never satisfied and the bound disappears. The only existing constraint in this mass range comes from measuring the long-range magnetic field of Jupiter  \cite{Jaeckel:2010ni} and is not very restrictive (see Fig.~\ref{fig_bound}). Existence of this constraint indicates that the plasma frequency is not identical to the photon mass. In particular, plasma frequency does not lead to screening of long-range static magnetic fields allowing to establish the bound.
 
 It is not a surprise that constraining ultra-light photons is hard. Indeed, 
 in the limit of zero mass the Lagrangian 
 (\ref{Lagr}) describes just the conventional electrodynamics with \[B_\mu={A_\mu+\varepsilon \tilde{A}_\mu\over (1+\varepsilon^2)^{1/2}}\] being the physical photon field,
  plus an additional massless vector field which is completely decoupled. In this case  $\varepsilon$ is not a physical parameter and no constraint on its value is possible\footnote{Of course, there is also no room for the hidden photon dark matter.}. This demonstrates that extra care is required when discussing the ultra-light regime. In fact, additional rather tight constraints in the mass range $10^{-24}\mbox{\rm~eV}\lesssim
 m\lesssim 10^{-18}\mbox{\rm~eV}$ were claimed in literature (from the drift of fine structure constant induced through
 the Zeeman effect \cite{Nelson:2011sf}, and, most recently, from the Voyager magnetometric survey \cite{Pignol:2015bma}).
As we explain below, these constraints do not apply, because cosmic plasma shielding, neglected in 
\cite{Nelson:2011sf,Pignol:2015bma}, invalidates both effects.

The principal goal of the present note is to improve on this situation and to push the bounds further into the ultra-light regime 
$m\ll 10^{-14}$~eV.  Not surprisingly, this requires  a further careful look into plasma effects.
The main idea is very simple. Hidden photon dark matter produces an oscillating electric field in the interstellar medium, which in turn induces electric current in its ionized component. Galactic plasma is not a perfect conductor, so the current dissipates resulting in a heat transfer from the hidden photon dark matter into ionized interstellar medium.
 By imposing that the rate of this heat transfer is smaller than the observed cooling rate from the dominant cooling mechanism for certain regions of the interstellar medium, we obtain constraints on the kinetic mixing coefficient $\varepsilon$. The resulting constraint in presented in Fig.~\ref{fig_bound}. It is somewhat stronger than the existing bound for $m\sim10^{-11}$~eV. However, its most notable feature is that it extends deep into the ultra-light mass regime, all the way down to $m\sim 10^{-20}$~eV.
 
 It is worth noting that, as discussed in \cite{Graham:2015rva}, hidden photon dark matter in the ultra-light mass range cannot be produced by inflationary perturbations due to isocurvature bounds, and the minimal misalignment mechanism is not efficient enough. However, there appears to be no reasons to exclude that a more general misalignment mechanism may be efficient. So it is worthwhile to study direct observational bounds in this parameter range.
 
 \section{Plasma equations with hidden photon}
 To derive this result let us study the dynamics of the photon and hidden photon fields in the interstellar medium, which we model as a non-relativistic plasma. Then the motion of electrons and ions is described by the momentum conservation equations (see, e.g., \cite{chen1984plasma})

\begin{equation}
m_e\frac{\partial \vec{u}_e}{\partial t} + m_e\left(\vec{u}_e\cdot \vec{\nabla}\right)\vec{u}_e = - {e\over (1+\varepsilon^2)^{1/2}}\left(\vec{E}+\varepsilon\vec{\tilde{E}} + \vec{u}_e\times(\vec{B}+\varepsilon\vec{\tilde{B}})\right)
- m_e\nu\left(\vec{u}_e - \vec{u}_i\right)
\label{eq_Langevin}
\end{equation}

\begin{equation}
m_i\frac{\partial \vec{u}_i}{\partial t} + m_i\left(\vec{u}_i\cdot \vec{\nabla}\right)\vec{u}_i = {e\over (1+\varepsilon^2)^{1/2}}\left(\vec{E}+\varepsilon\vec{\tilde{E}} + \vec{u}_e\times(\vec{B}+\varepsilon\vec{\tilde{B}})\right) - m_e\nu\left(\vec{u}_i - \vec{u}_e\right)
\label{eq_Langevin_ion}
\end{equation}
%
%
%
where $\vec{u}_i$ and $\vec{u}_e$ are the velocity fields of the electrons and ions respectively.
The first  term on the r.h.s. of both equations represent the Lorentz forces experienced by charged particles due to  the conventional  electromagnetic fields and due to the hidden one.
The last term on the r.h.s. of both equations accounts for the friction due to electron-ion collisions. The frequency of electron-ion collisions $\nu$ is given by
\begin{equation}
\nu = {4\sqrt{2\pi}\alpha^2n_e\over 3m_e^{1/2}T_e^{3/2}}
\log\Lambda_C\approx 3\cdot 10^{-21}\mbox{\rm ~eV}
\l{n_e\over 0.04\mbox{\rm ~cm}^{-3}}\r
\l{ 8000 \mbox{\rm ~K}\over T_e}\r^{3/2} \;,
\end{equation}
where  $n_e$ is the electron density, $T_e$ is the electron temperature
and the Coulomb logarithm is given by
\[
\log\Lambda_C=\log \l{4\pi T_e^3\over\alpha^3n_e}\r^{1/2}\approx 25\;,
\]
for typical parameters of interstellar medium ($n_e\sim 10^{-2}$~cm$^{-3}$, $T_e\sim 8000$~K) and we neglected ions with charge larger than 1.

Eqs.~(\ref{eq_Langevin}) are complemented by the electromagnetic and Proca field equations
\begin{gather}
\Box A^\mu = {e\over (1+\varepsilon^2)^{1/2}}J^\mu    \nonumber\\
\Box \tilde{A}^\mu + m^2\tilde{A}^\mu =  {e\varepsilon\over (1+\varepsilon^2)^{1/2}} J^\mu\nonumber
\end{gather}
\noindent and the fields satisfy \[\partial_\mu A^\mu = \partial_\mu \tilde{A}^\mu = 0\;,\] 
as a consequence of  the Landau gauge choice for the electromagnetic field and the Proca constraint for the hidden photon field. Finally the current is written in terms of the electron/ion density and velocity fields as
\begin{equation}
J^\mu = \left(\rho, \vec{J}\right) = \left(n_i,n_i \vec{u_i}\right)-\left(n_e,n_e \vec{u_e}\right) 
\end{equation}
To study wave propagation in the cold plasma  we linearize the above system of equations around a configuration with zero photon and hidden photon fields, zero electron velocity and constant electron and ion densities $n_e$ and $n_i$. That is, we linearize the system by considering the (constant) densities to be zero-order quantities and the velocity fields and electric and hidden electric fields to be first-order quantities. The magnetic fields can be neglected at the first order. The linearized momentum conservation equations take the form 
\begin{gather}
m_e\frac{\partial \vec{u}_e}{\partial t} = - {e\over (1+\varepsilon^2)^{1/2}}(\vec{E}  +\varepsilon\vec{\tilde{E}} ) - m_e\nu\left(\vec{u}_e - \vec{u}_i\right)
\label{eq_Langevin_lin}\\
m_i\frac{\partial \vec{u}_i}{\partial t}  = {e\over (1+\varepsilon^2)^{1/2}}(\vec{E}  +\varepsilon\vec{\tilde{E}} )- m_e\nu\left(\vec{u}_i - \vec{u}_e\right).
\label{eq_Langevin_ion_lin}
\end{gather}
 At this order the total momentum of the system is conserved at each point,  
 so that in the rest frame we find
  \[
  m_e\vec{u}_e + m_i\vec{u}_i = 0\]
  at all times.
   Then at the leading order in $m_e/ m_i$ we may set the ion velocity to zero,
   Eq.~(\ref{eq_Langevin_ion_lin}) holds trivially and we are left  with Eq.~(\ref{eq_Langevin_lin}) at $u_i=0$.
\section{Plasma shielding}
After switching to the Fourier modes for all first-order quantities,
\begin{equation}
A(\vec{x},t) = \frac{1}{(2\pi)^{3/2}}\int d^3k d\omega A(\vec{k},\omega)e^{i\vec{x}.\vec{k} - i\omega t}
\end{equation}
we may solve from Eq.(\ref{eq_Langevin_lin}) for the electron velocity in terms of the electric fields as
\begin{equation}
\vec{u}_e(\vec{k},\omega) = \frac{-e}{m_e\left(\nu-i\omega\right)}{\vec{E} + \varepsilon\vec{\tilde{E}}\over (1+\varepsilon^2)^{1/2}}.
\end{equation}

Plugging this into the electromagnetic equations for the photon and hidden photon fields and decomposing the spatial parts of the vector fields into transverse and longitudinal parts
\be
A_i(\vec{k},\omega) = A_L \frac{k_i}{\vec{k}^2} + A_{Ti}\,,\;\;\;\;\; A_{Ti}k_i = 0
\ee
we find 
\begin{equation}
\left\lbrace \vec{k}^2 - \omega^2 +(1+\varepsilon^2)^{-1}
\left(\begin{array}{cc}
\Omega_p^2 & \Omega_p^2\varepsilon \\
\Omega_p^2\varepsilon & \Omega_p^2\varepsilon^2 + m^2(1+\varepsilon^2)
\end{array}
\right)
\right\rbrace
\left(
\begin{array}{c}
A_{T}\\
\tilde{A}_{T}
\end{array}
\right) = 0
\label{matrix_transverse}
\end{equation}
for the transverse components of the vector fields, and 
\begin{equation}
\left\lbrace  - \omega^2 +
(1+\varepsilon^2)^{-1}\left(\begin{array}{cc}
\Omega_p^2 &\Omega_p^2\varepsilon \\
\Omega_p^2\varepsilon & \Omega_p^2\varepsilon^2 + \frac{m^2(1+\varepsilon^2)}{1 - \frac{\vec{k}^2}{\omega^2}}
\end{array}
\right)
\right\rbrace
\left(
\begin{array}{c}
A_{L}\\
\tilde{A}_{L}
\end{array}
\right) = 0
\label{matrix_longitudinal}
\end{equation}
 for the  longitudinal component,
 where
 \[
 \Omega_p^2\equiv {\omega_p^2\over 1+{i \nu\over\omega}}\;.
 \]

We see that the presence of matter affects the vacuum mixing  of the two fields $A$ and $\tilde{A}$, as anticipated. We may now diagonalize the previous set of equations to obtain the two propagating modes and their respective dispersion relations for the transverse and longitudinal parts of the fields.

 Let us first discuss the structure of the propagating modes. 
For the purpose of the dark matter discussion we are interested in the non-relativistic regime $k\ll \omega$.
Then the transverse and longitudinal equations approach the same form and the propagating modes 
 correspond to the eigen-modes of the matrix 
\begin{equation}
M^2 = (1+\varepsilon^2)^{-1}\left(\begin{array}{cc}
\Omega_p^2 & \Omega_p^2\varepsilon \\
\Omega_p^2\varepsilon & \Omega_p^2\varepsilon^2 + m^2(1+\varepsilon^2)
\end{array}
\right)\;.
\end{equation}
At $m\ll\omega_p$ the two eigenvalues of the matrix $M^2$ take the form
\begin{gather}
\omega_v^2=\Omega_p^2\l1+{\cal O}\l{m^2\over \Omega_p^2}\r\r\\
\omega_h^2= {m^2\over 1+\varepsilon^2}\l1+{\cal O}\l{m^2\over \Omega_p^2}\r\r\;.
\end{gather} 
Ultra-light dark matter is comprised of the second (lighter) mode $A_{DM}$. The corresponding $A$ and $\tilde{A}$ components take the form
\begin{gather}
A=-A_{DM}
{\varepsilon\over (1+\varepsilon^2)^{1/2}}
\l1+{m^2\over \Omega_p^2(1+\varepsilon^2)}\r+{\cal O}\l{m^4\over \Omega_p^4}\r\\
\tilde{A}=A_{DM}{1\over (1+\varepsilon^2)^{1/2}}
\l1-{m^2\varepsilon^2\over \Omega_p^2(1+\varepsilon^2)}\r+{\cal O}\l{m^4\over \Omega_p^4}\r\;.
\end{gather}
To see the plasma shielding effect let us calculate the corresponding observable electromagnetic field given by 
\be
\label{suppr}
A_{\text{obs}} ={ A + \varepsilon\tilde{A}\over (1+\varepsilon^2)^{1/2}} 
\approx -{\varepsilon\over 1+\varepsilon^2}\frac{m^2}{\Omega_p^2}A_{DM}. 
\ee
We see that on top of a vacuum mixing suppression present at small $\varepsilon$, interactions of ultra-light hidden photon dark matter with the SM matter are suppressed by an additional
factor  
\be
{m^2\over \Omega_p^2}\approx
\left\{\begin{array}{l}
{ m^2\over \omega_p^2}\;\;\text{ for } \nu\ll m\ll\omega_p\\
{i\nu m\over  \omega_p^2}\;\;\text{ for } m\ll\nu
\end{array}
\right.
\ee 
It is straightforward to understand the physical origin of this suppression. Given that the plasma frequency is much higher than the hidden photon mass the light mode should be highly aligned with a linear combination
of $A$ and $\tilde{A}$, which does not interact with plasma,  in order to stay light.
As a consequence of this suppression, direct detection of ultra-light photons becomes very challenging.
For instance, strong bounds on the value of $\varepsilon$ in the mass range $10^{-23}$~eV$<m<10^{-18}$~eV were claimed recently in \cite{Pignol:2015bma}, based on the precision magnetometry performed on board of the Voyager space probes. However, including the plasma shielding effect derived here and neglected in \cite{Pignol:2015bma}, reduces the expected signal at least by a factor of   $10^{-12}$ (for $m=10^{-18}$~eV and  taking  $n_e\sim10^{-3}$~cm$^{-3}$, which is appropriate for the heliosheath region\cite{helisheath}) completely eliminating the constraint.
 
 The plasma shielding suppression cannot be directly applied to the Earth based experiments, in particular to the constraint based on the Zeeman effect \cite{Nelson:2011sf}, and  to the SQUID based setups proposed in 
 \cite{Pignol:2015bma}, because these experiments are performed in the dielectric environment of the Earth atmosphere.
However, a strong signal suppression of a very similar origin applies in this case as well. Indeed, here we encounter the situation analyzed  in detail in \cite{Chaudhuri:2014dla}. Namely, effectively the experiment is performed in a vacuum (dielectric) cavity surrounded by a conducting medium. As demonstrated in \cite{Chaudhuri:2014dla}, for small photon masses the electric field inside the cavity is suppressed then by an additional factor of 
$\max \{{\cal O}(m^2R^2),{\cal O}(m R v_{DM})\}$ and the magnetic field is suppressed by an additional factor of ${\cal O}(mR)$, where $v_{DM}\sim 10^{-3}$ is the dark matter velocity and $R$ is the size of a cavity.
For the mass range $m\lesssim 10^{-19}$~eV considered in \cite{Nelson:2011sf} the bound disappears even if one estimates the size of a cavity to be $R\sim R_{Earth}\sim 6000$~km. Given that the Earth soil has very decent conducting properties it is likely that the actual suppression is much stronger and corresponds to $R\sim 60$~km, which is the height of the Earth ionosphere.  A proper accounting for this suppression is definitely required to evaluate the feasibility of the SQUID magnetometry proposal of \cite{Pignol:2015bma}.

To conclude this discussion it is worth noting that the plasma shielding effect does not affect the bound based on the study of the static magnetic field of the Jupiter. Indeed, this bound is based on the study of a kinematical regime 
$k\gg\omega$, which is very different from the non-relativistic regime discussed so far.  
Plasma does not screen static magnetic fields and the Jupiter bound applies.
\section{Heating up the Galaxy}
To summarize the above discussion,
the plasma shielding effect  leaves hidden photon dark matter practically unconstrained for the mass range\footnote{Barring the Jupiter constraint for the heavy end of this mass interval, which is rather mild. See also section~\ref{sec:last} for a brief discussion of the superradiance constraint. For masses much lighter than $10^{-23}$~eV, the hidden photon is not a viable dark matter candidate, because its de Broigle wavelength is longer than the observed small scale structures.} $10^{-23}\;\text{eV}<m<10^{-14}$~eV. As we will see now, a further look into plasma effects allows nevertheless to establish a quite stringent constraint over a sizable part of this mass range. The idea behind this constraint is that an oscillating electric field induced by hidden photon dark matter should dissipate as a consequence of non-zero resistivity of the interstellar plasma. The resulting  heat transfer rate from dark matter into the interstellar medium should not be too fast.

 To calculate the heat transfer rate we need to find the imaginary part of the frequency corresponding to the dark matter mode. As before, we restrict to the non-relativistic limit, so that the eigenfrequency equations are the same for 
 the transverse and longitudinal modes. Even though we are mostly interested in the ultra-light regime, $m\ll \omega_p$, for completeness we consider also the opposite regime, $m\gg \omega_p$.
 The dispersion relations of non-relativistic plasma waves  
are determined from the following equation, resulting from (\ref{matrix_transverse}) (or, equivalently, from (\ref{matrix_longitudinal})),
\be
\omega^2 = {1\over 2}\l m^2 +\frac{\omega_p^2}{\left(1+ i\frac{\nu}{\omega}\right)}\\
\pm \sqrt{\left(m^2 + \frac{\omega_p^2}{\left(1+ i\frac{\nu}{\omega}\right)}\right)^2 - \frac{4\omega_p^2 m^2}{\l1+\varepsilon^2\r\left(1 + \frac{i\nu}{\omega}\right)}}\r.
\ee
As before,  the two branches correspond to the conventional plasma oscillations and to the dark matter mode.
They exchange their roles 
 for $m$ above and below $\omega_p$. Namely, for $m \ll \omega_p$ it is the negative branch that behaves like the hidden photon, whereas for $m \gg \omega_p$ it is the other way around. Taking this into account and decomposing the frequency into its real and imaginary parts $\omega = \omega_h + i\gamma_h$ we have in the two limits
\begin{equation}
\omega_h^2  = \left\lbrace\begin{array}{ll}
{m^2\over 1+\varepsilon^2}\text{  for $m \ll \omega_p$}\\
m^2 \text{  for $m \gg \omega_p$}
\end{array}\right. 
\end{equation}  
and
\begin{equation}
\gamma_h = \left\lbrace\begin{array}{ll}
-\nu\frac{m^2}{2\omega_p^2}\frac{\varepsilon^2}{1+\varepsilon^2}  \text{   for $m \ll \omega_p$}\\
-\nu\frac{\omega_p^2}{2m^2}\frac{\varepsilon^2}{1+\varepsilon^2} \text{  for $m \gg \omega_p$}\;.
\end{array}\right. 
\end{equation}
%
%
%
From this last expression we obtain the rate of the energy loss for the dark matter hidden photon field
\[
Q = 2\arrowvert\gamma\arrowvert \rho_{h}\;,
\]
where $\rho_h$ is the dark photon energy density.
This translates into
\begin{equation}
Q \approx
2\cdot 10^{-9}
\varepsilon^2
{
\text{erg}\over\text{s}\,\text{cm}^3
}
\l
{m
\over
7.4\cdot 10^{-12}\text{ eV}}
\r^2
\l
{\rho_{h}\over 0.3 \text{ GeV}\,\text{cm}^{-3}}
\r
\l
{8000\text{ K}\over T_e}
\r^{3/2}
	\label{lighteq_heat}
\end{equation}
in the light mass regime ($m \ll \omega_p$) and into 
\begin{equation}
Q \approx
2\cdot 10^{-9}
\varepsilon^2
{
\text{erg}\over\text{s}\,\text{cm}^3
}
\l
{7.4\cdot 10^{-12}\,\text{eV}}
\over
m
\r^2
\l
{\rho_{h}\over 0.3 \text{ GeV}\,\text{cm}^{-3}}
\r
\l
{8000\,\text{K}\over T_e}
\r^{3/2}
\l
{
n_e\over 0.04\text{ cm}^{-3}
}
\r^2
	\label{heavyeq_heat}
\end{equation}
in the heavy mass regime ($m \gg \omega_p$), where in both cases we assumed $\varepsilon^2\ll 1$. Interestingly, in the light mass regime the energy loss rate is independent of the electron density. This comes out as a
result of the cancelation between the two competing effects---decreasing the electron density suppresses the collision rate $\nu$, but at the same time the plasma shielding effect is less pronounced in the dilute plasma.

This energy gets dumped into a thermal motion of plasma electrons. 
The current state of the interstellar medium is maintained through a delicate balance between various cooling and heating mechanisms (see, e.g., \cite{2011piim.book.....D,lequeux2005interstellar} for an introduction), which will be destabilized if the heat transfer rate from the dark matter photons is too fast. Unfortunately, there is no direct measurement of the heating rate, although theoretical expectations for a number of key processes are available.
On the other hand, there are more observational handles on the cooling rate (and it is encouraging that the results match well theoretical expectations for the heating rates). So to obtain a bound on the hidden photon parameters, we impose that the rate given by (\ref{lighteq_heat}), (\ref{heavyeq_heat}) is smaller than the observed cooling rate for various regions of the interstellar medium.

 One way to arrive at the bound is to consider the warm ionized medium, a dilute phase of the interstellar medium of temperature around $8000 \text{ K}$ \cite{lequeux2005interstellar}, extending about the galactic midplane with a thickness of more than $1\text{ kpc}$, filling a volume fraction $f \sim 0.4\div 0.2$ \cite{haffner2009warm}.  For this phase, the cooling is dominated by the inelastic collisions of electrons and hydrogen atoms with singly ionized carbon atoms $C_{II}$ \cite{lequeux2005interstellar}.  The ground state of $C_{II}$ is split into two fine structure levels, and inelastic collisions can excite transitions from the $^2P_{1/2}$ level to the $^2P_{3/2}$ level (excited atoms being labeled $C_{II}^{*}$). Subsequent de-excitation of the excited fine structure level yields the far-infrared emission in the $157.7 \text{ }\mu m$ line.

The rate of cooling due to this mechanism can be deduced from the abundance of $C_{II}^{*}$  in the warm ionized medium. For different regions of the interstellar medium this abundance can be approximated from the integrated column density of $C_{II}^{*}$ along different lines of sight. The latter  can be in turn obtained from the $C_{II}^{*}$ absorption lines in the far ultraviolet given by transitions originating in the $^2P_{3/2}$ level.

 The procedure outlined above has been carried out by several authors \cite{matsuhara1997observations,pottasch1979determination,LehnerCII} to obtain the cooling rate from $C_{II}$ emission into the infrared. In \cite{LehnerCII}, measurements of the $1037.018$~A and $1335.708 $~A far UV absorption lines of 43 objects at high galactic latitude ($b \gtrsim 30^{\circ} $) are used to obtain the cooling rates per hydrogen atom for the clouds along the corresponding lines of sight. The gas probed along these lines of sight is claimed to be at least partialy in the warm ionized phase. The results for the cooling rates vary between an average of $\sim 10^{-26} n_H\text{ ergs/s}$ for directions corresponding to low velocity clouds and a smaller value of around $\sim 10^{-27} n_H\text{ ergs/s}$ for higher velocity clouds, where $n_H$ is the density of Hydrogen atoms. The latter 
  is itself expected to be  in the $0.1\div 1 \text{ cm}^{-3} $ range \cite{ferriere2001interstellar,dong2011halpha} (and may also be estimated from the data presented in  \cite{LehnerCII} from the available  values of the electron density and ionization fraction).
  
 If the heating rate due to hidden photons were significantly larger than the measured cooling rate of the interstellar medium,  this would change thermal dynamics of the warm ionized medium (WIM) with a timescale of order
   \begin{equation}
  \Delta t = T/\dot{T} \sim 10^{14}\mbox{\rm ~s} \left(\frac{\varepsilon_0}{\varepsilon}\right)^2
  \label{timescale_coolingchange}
  \end{equation}
  where in the last estimate we used $\dot{T} \sim Q/n_H\sim 10^{-26}\text{ ergs/s}$, $T\sim 8000$~K and $\varepsilon_0(m)$ is the value of $\varepsilon$ for which the heating and cooling rates are equal. For regions considered in \cite{LehnerCII}
  the cooling rate $\Lambda$ is mostly due to electron collisions with the carbon ions and it is given by
  \begin{equation}
  \label{cool}
\Lambda \simeq 10^{-19} 
n_e n_C
e^{-91.2\text{ K}/T}\left(\frac{T}{1 \text{K}}\right)^{-1/2}\text{ ergs s}^{-1}\text{ cm}^{3}\;,
\end{equation}
where $n_C$ is the carbon number density.
The regions  considered in \cite{LehnerCII} have relatively high temperature $T \approx 8000 \text{ K}$. Then the temperature dependence of the cooling rate is dominated by the $T^{-1/2}$ prefactor in 
  (\ref{cool}) and  we find that
\[ T/\dot{T} \sim \Lambda/\dot{\Lambda}\]
 and the timescale given by (\ref{timescale_coolingchange}) describes also the time necessary for the cooling rate to change significantly. The characteristic time scale (\ref{timescale_coolingchange}) is very fast on cosmological scales, which leads to the conclusion that the heating rate cannot exceed the observed cooling rate.
 
  A similar bound can be deduced from inspecting the  dependence of the cooling rate of the WIM on the temperature \cite{lequeux2005interstellar}.
   One finds that for an ionization fraction around $n_e/n_H \sim 0.1$ and $n_H \sim 0.1 \text{ cm}^{-3}$, the heating rate higher than about  $10^{-26} \text{erg s cm}^{-3}$ would necessarily bring the system towards thermal equilibrium in a regime where $T \sim 10^4 \text{ K}$, and the cooling rate is dominated by Lyman $\alpha$ emission. 
     
 Imposing the condition that the heating rates  (\ref{lighteq_heat}) and (\ref{heavyeq_heat}) are smaller than the observed cooling rates we find the bound presented  in Fig.~\ref{fig_bound}. We accounted for the uncertainty in the value of $n_H$ and other parameters, by presenting the bound for several choices of the parameters, as explained in the caption.
 
\section{Conclusions}
\label{sec:last}
To summarize, in this note we elaborated on plasma effects in the ultra-light hidden photon  dark matter model. On one side, we pointed out that the previously neglected  plasma shielding effect makes it quite challenging to directly detect ultra-light hidden photon dark matter. On the other hand we derived a quite stringent bound on the mixing parameter of ultra-light hidden photon dark matter, resulting from the measured cooling rates in the Galactic warm ionized medium.
It appears plausible that our bound may be somewhat sharpened by a dedicated analysis of the cooling rate in specific regions or in a different environment. Given that the heat transfer rate from dark matter to the interstellar medium is independent of the electron density, one may consider, for example, also  warm HI regions. A brief look at the corresponding
 typical cooling rates and ionization fractions presented in \cite{2011piim.book.....D} suggests that indeed this results in a comparable  bound.  It would be interesting to see whether the bound can be further improved by considering the intergalactic medium either in the current epoch or at higher redshifts. The heat transfer rate in this case gets suppressed compared to what one finds in the Galaxy due to smaller dark matter density. However, this  may well be compensated by an even stronger decrease in the cooling rate, which is proportional to $n_e^2$. We leave this kind of analysis, requiring more detailed study of astronomical data, for the future.
 
Our bound becomes weaker as the photon mass decreases, and disappears at masses around $10^{-20}$~eV. It will be interesting to explore other ways to constrain ultra-light hidden photon dark matter with 
these tiny masses. We are aware of two ideas how this can be achieved. 
First, similarly to ultra-light axions  \cite{Arvanitaki:2009fg,Arvanitaki:2010sy}, also ultra-light vectors in the mass range $\sim 10^{-22}\div 10^{-10}$~eV may be constrained by measuring spins of astrophysical black holes, through the effect of black hole superradiance. Existing spin measurements for stellar mass black holes constrain the mass range $\sim 10^{-12}\div10^{-11}$~eV \cite{Arvanitaki:2014wva}. It is unclear whether the present spin determinations for supermassive black holes are robust enough to use them for establishing the bound. By taking them at the face value, one may exclude the mass range $\sim 10^{-20}\div10^{-16}$~eV \cite{Pani:2012vp,Arvanitaki:2014wva}. Note, that the superradiance bound does not rely on a particle being a significant part of dark matter, and applies at $\varepsilon=0$. On the other hand,
matter effects may inhibit the superradiance instability at sufficiently large values of $\varepsilon$. To the best of our knowledge, these were not studied for hidden photons, so at the moment it is unclear what is the largest value of $\varepsilon$ such that the bound applies.
 
 In addition, it was proposed recently in \cite{Khmelnitsky:2013lxt} that ultra-light scalar dark matter in the mass range $10^{-23}\div10^{-22}$~eV produces an oscillating gravitational potential, which may be observed with pulsar timing arrays. It is natural to expect that this also applies to the ultra-light photons in the same mass range. It will be interesting to study whether the two models can be distinguished observationally.

We would like to thank Andrei Gruzinov for very helpful discussions. We would also like to thank Ely Kovetz for pointing out a problem with Fig. \ref*{fig_bound} in a previous version of the paper. This work was supported in part by the NSF CAREER award PHY-1352119.
\bibliographystyle{utphys}
\bibliography{biblio}

\end{document}